\documentclass[12pt]{article}
\usepackage{amsmath}

\begin{document}
 
\title{Hydrodynamic reductions of the heavenly equation}
\author{E.V.Ferapontov \& M.V.Pavlov \\
Department of Mathematical Sciences \\
Loughborough University \\
Loughborough, Leicestershire LE11 3TU \\
United Kingdom \\
e-mail: \texttt{E.V.Ferapontov@lboro.ac.uk}}
\date{}
\maketitle
 
\begin{abstract}

We demonstrate that Pleba\'nski's first heavenly equation decouples in infinitely many ways into a triple of commuting $(1+1)$-dimensional systems of hydrodynamic type which satisfy the Egorov property. Solving these systems by the generalized hodograph method, one can construct exact solutions of the heavenly equation parametrized by arbitrary functions of a single variable. 
We discuss explicit examples of hydrodynamic reductions  associated with the  equations of one-dimensional nonlinear elasticity, linearly degenerate systems and the equations of associativity.
 
 
\end{abstract}

\newpage
 
\section{Introduction}

The Pleba\'nski first heavenly equation,
\begin{equation}
\Omega_{xy}\Omega_{zt}-\Omega_{xt}\Omega_{zy}=1,
\label{P1}
\end{equation}
governs K\"ahler potentials of  the 4-dimensional metrics
$$
ds^2=2\Omega_{xy}dxdy+2\Omega_{zt}dzdt+2\Omega_{xt}dxdt+2\Omega_{zy}dzdy,
$$
which are simultaneously Ricci-flat and self-dual \cite{Penrose, Plebanski, Ko}.

This equation was thoroughly investigated during the past three decades using various technics.
It's integrability via the twistor construction was established  in \cite{Penrose}.
The symmetry properties and conservation laws of the heavenly equation  were studied in  \cite{Winternitz, Strachan1, Strachan2, Nutku, Boyer2, Alfinito}. We refer to the review \cite{Boyer1} for further references.

In this paper we propose an alternative approach to the construction of exact solutions of
the PDE (\ref{P1}), based on its decomposition into a family of commuting (1+1)-dimensional systems of hydrodynamic type. This method was successfully applied to the dispersionless 
KP  equation  
in a series of publications \cite{Gibb94, GibTsa96,
GibTsa99, Lei, Ma}, and subsequently extended to the Boyer-Finley equation in
\cite{MaAl, GuMaAl, Fer}. In principle, it applies to any multidimensional dispersionless
`integrable' system. Although, written in the form (\ref{P1}), the heavenly equation is not quasilinear, one can introduce new  variables
\begin{equation}
\Omega_{xy}=a, ~~ \Omega_{zt}=b, ~~ \Omega_{xt}=p, ~~ \Omega_{zy}=q, 
\label{P2}
\end{equation}
transforming (\ref{P1}) into the quasilinear form
\begin{equation}
a_t=p_y, ~~ a_z=q_x, ~~ b_x=p_z, ~~ b_y=q_t, ~~~~ ab-pq=1.
\label{P3}
\end{equation}
Hydrodynamic reductions are sought in the form
\begin{equation}
a=a(R^1, ..., R^n), ~~ b=b(R^1, ..., R^n), ~~ p=p(R^1, ..., R^n), ~~ q=q(R^1, ..., R^n), 
\label{abpq}
\end{equation}
where the dependence of the `Riemann invariants'  $R^1, ..., R^n$ on $x, y, z, t$ is governed by a 
triple of commuting hydrodynamic type systems 
\begin{equation}
R^i_t=v^i(R)\ R^i_x, ~~~~ R^i_y=w^i(R)\ R^i_x, ~~~~ R^i_z=\mu^i(R)\ R^i_x; 
\label{nP1}
\end{equation}
(notice that the number $n$ of Riemann invariants is allowed to be arbitrary!)
We recall, see \cite{Tsarev}, that the requirement of the commutativity of the flows (\ref{nP1})
is equivalent the following restrictions on their characteristic speeds:
$$
\frac{\partial_jv^i}{v^j-v^i}=\frac{\partial_jw^i}{w^j-w^i}=\frac{\partial_j\mu^i}{\mu^j-\mu^i}, ~~~ i\ne j,
~~~ \partial_j=\partial/\partial_{ R^j}.
$$ 
Substituting the ansatz (\ref{abpq}) into the equations (\ref{P3}) and using (\ref{nP1}), one readily arrives at a system of PDEs for $a, b, p, q, v^i, w^i, \mu^i$ as functions of
$R^1, ..., R^n$, see equations (\ref{nP2}) in Sect. 2. Although overdetermined, this system is compatible, with the  general solution depending on $3n$ arbitrary functions of a single variable.
All hydrodynamic reductions of the form (\ref{nP1}) are automatically semi-Hamiltonian
\cite{Tsarev}, that is, their characteristic speeds satisfy the additional relations
\begin{equation}
\partial_k\frac{\partial_jv^i}{v^j-v^i}=\partial_j\frac{\partial_kv^i}{v^k-v^i}, ~~~ i\ne j\ne k,
\label{semi}
\end{equation}
(the same constraints holding for $w^i$ and $\mu^i$). Relations (\ref{semi}) imply the existence of the diagonal metric $\sum g_{ii}(dR^i)^2$ such that
\begin{equation}
\frac{\partial_jv^i}{v^j-v^i}=\partial_j\ln \sqrt{g_{ii}}, ~~~~ i\ne j.
\label{metric}
\end{equation}
This metric  is  clearly the same for the whole family of commuting flows (\ref{nP1}),
$$
\frac{\partial_jv^i}{v^j-v^i}=\frac{\partial_jw^i}{w^j-w^i}=\frac{\partial_j\mu^i}{\mu^j-\mu^i}=
\partial_j\ln \sqrt{g_{ii}};
$$
it plays an important role in the Hamiltonian formulation of the corresponding hierarchy \cite{Dub, Tsarev}. The generalized hodograph method proposed by Tsarev 
\cite{Tsarev}, provides the general solution of the system (\ref{nP1}) by  the implicit formula
\begin{equation}
\lambda^i(R)=x+v^i(R)\ t+w^i(R)\ y+\mu^i(R) \ z, ~~~ i=1, ..., n,
\label{hod}
\end{equation}
where $\lambda^i(R)$ are the characteristic speeds of the general flow commuting with (\ref{nP2}), that is, the general solution of the
linear system
\begin{equation}
\frac{\partial_j\lambda^i}{\lambda^j-\lambda^i}=\partial_j\ln \sqrt{g_{ii}}, ~~~~ i\ne j.
\label{com}
\end{equation}
Ultimately, the method of hydrodynamic reductions consists of the following steps:

--- find $a,\ b,\ p,\ q, v^i, w^i, \mu^i$ as functions of $R^1, ..., R^n$ by solving the system (\ref{nP2});

--- solve the linear system (\ref{com});

--- express $R^1, ..., R^n$ as functions of $x, y, z, t$ from the  implicit  relations (\ref{hod});

--- reconstruct $\Omega$ by virtue of (\ref{P2}).

In Sect. 2 we derive the involutive system of PDEs governing $n$-component  reductions of the heavenly equation. We demonstrate, in particular, that all such reductions satisfy the Egorov property, that is, the diagonal metric $g_{ii}$ defined by (\ref{metric}) possesses a potential:
$$
g_{ii}=\partial_i\psi.
$$
Diagonal systems satisfying the Egorov property play a particularly important role in applications:
examples thereof inlude isentropic gas dynamics, chemical kinetics, equations obtained by the Whitham averaging,
hydrodynamic reductions of the dispersionless KP and Boyer-Finley equations, etc.
Egorov's (potential) metrics and the corresponding  hydrodynamic hierarchies naturally arise in the context  of 2D topological field theories \cite{Dubr}. Multi-Hamiltonian Egorov's systems
were recently classified in \cite{Pavlov}.

One-component reductions ($n=1$) of the heavenly equation are described in Sect. 3.
Examples of  two-component reductions associated with the equations of nonlinear elasticity
are provided in Sect. 4. Some multi-component examples are discussed in Sect. 5.

In this paper we concentrate on mathematical aspects of the method.  The geometry of
the examples constructed requires a special investigation.

{\bf Remark 1.} More general hydrodynamic reductions can be sought in the form
$a=a(u^1, ..., u^n)$, $ b=b(u^1, ..., u^n), $ $ p=p(u^1, ..., u^n), $ $ q=q(u^1, ..., u^n) $, where the dependence of
$u^1, ..., u^n$ on $x, y, z, t$ is governed by a 
triple of commuting hydrodynamic type systems 
$$
u^i_t=v^i_j(u)\ u^j_x, ~~~~ u^i_y=w^i_j(u)\ u^j_x, ~~~~ u^i_z=\mu^i_j(u)\ u^j_x, 
$$
which are no longer assumed to be strictly hyperbolic or diagonalizable. Reductions of this type are presented in Example 3 (elliptic case) and Example 5 (parabolic case).

\section{Properties of $n$-component reductions}

Substituting  $a(R), \  b(R), \ p(R), \ q(R)$ into the equations (\ref{P3}) and using
(\ref{nP1}), one readily arrives at  the equations
\begin{equation}
\partial_iq=\mu^i\partial_ia, ~~~ \partial_ip=\frac{v^i}{w^i}\partial_ia, ~~~
\partial_ib=\frac{\mu^i v^i}{w^i}\partial_ia;
\label{partial}
\end{equation}
the differentiation of $ab-pq=1$ implies $a\partial_ib+b\partial_ia=p\partial_iq+q\partial_ip$, which, by virtue of (\ref{partial}), can be written in the form
\begin{equation}
v^i=w^i\frac{p\mu^i-b}{a\mu^i-q}.
\label{v}
\end{equation}
Substituting (\ref{v}) back into (\ref{partial}), one obtains
\begin{equation}
\partial_iq=\mu^i\partial_ia, ~~~
\partial_ip=\frac{p\mu^i-b}{a\mu^i-q}\partial_ia, ~~~
\partial_ib=\mu^i \frac{p\mu^i-b}{a\mu^i-q}\partial_ia,
\label{pb}
\end{equation}
A direct calculation of the compatibility conditions for equations (\ref{pb}) results in
the following over-determined system for $a$:
\begin{equation}
\begin{array}{c}
\partial_i\partial_ja=\frac{\partial_j\mu^i}{\mu^j-\mu^i}\partial_ia+\frac{\partial_i\mu^j}{\mu^i-\mu^j}\partial_ja, \\
\ \\
\partial_j\mu^i(a\mu^j-q)\partial_ia+\partial_i\mu^j(a\mu^i-q)\partial_ja=
(\mu^i-\mu^j)^2\partial_ia\partial_ja.
\end{array}
\label{a}
\end{equation}
The commutativity conditions,
\begin{equation}
\frac{\partial_jv^i}{v^j-v^i}=\frac{\partial_jw^i}{w^j-w^i}=\frac{\partial_j\mu^i}{\mu^j-\mu^i},
\label{b}
\end{equation}
with $v^i$ given by (\ref{v}),  imply 
\begin{equation}
\partial_j\mu^i=\frac{w^i\partial_ja(\mu^j-\mu^i)^2}{w^i(a\mu^j-q)-w^j(a\mu^i-q)}.
\label{c}
\end{equation}
Ultimately, combining (\ref{a}), (\ref{b}) and (\ref{c}), we arrive at the following equations for $a, q, w^i, \mu^i$:
\begin{equation}
\begin{array}{c}
\partial_iq=\mu^i\partial_ia, ~~~ \frac{\partial_jw^i}{w^j-w^i}=\frac{\partial_j\mu^i}{\mu^j-\mu^i}, \\
\ \\
\partial_i\partial_ja=\frac{\partial_ia\partial_ja (w^i+w^j)(\mu^j-\mu^i)}{w^i(a\mu^j-q)-w^j(a\mu^i-q)},  \\
\ \\
\partial_j\mu^i=\frac{w^i\partial_ja(\mu^j-\mu^i)^2}{w^i(a\mu^j-q)-w^j(a\mu^i-q)}, 
\label{nP2}
\end{array}
\end{equation}
$i\ne j$; notice that $p, b, v^i$ do not explicitly enter these equations. One can show by a direct calculation that the system (\ref{nP2}) is in involution, and its general solution depends on $3n$ arbitrary functions of a single variable.
For any solution of this system the characteristic speeds $v^i$ of the first commuting flow 
are given by the formula (\ref{v})
where the functions $p(R)$ and $b(R)$ can be recovered from the equations (\ref{pb}),
which are compatible by  construction. Moreover, equations (\ref{nP2}) and (\ref{pb}) are consistent with the algebraic restriction
$ab-pq=1$. Thus, the construction of $n$-component hydrodynamic reductions of the heavenly equation is reduced to the 
solution of the system (\ref{nP2}). 

{\bf Remark 2.} Introducing a potential $\psi$ by the formula
$\partial_i\psi=\partial_i a/w^i$ (the compatibility is guaranteed by (\ref{nP2})), one can  verify the identity
$$
\frac{\partial_j\mu^i}{\mu^j-\mu^i}=\frac{\partial_i\partial_j \psi}{2\partial_i\psi}, ~~~ i\ne j,
$$
implying that reductions (\ref{nP1}) satisfy the Egorov property. The corresponding Egorov metric is
$\sum \partial_i\psi (dR^i)^2$.

{\bf Remark 3.} We have demonstrated that the system
$$
R^i_z=\mu^i(R)\ R^i_x,
$$
which is the third commuting flow in (\ref{nP1}), satisfies the Egorov property and possesses a pair of conservation laws 
$$
a(R)_z=q(R)_x, ~~ p(R)_z=b(R)_x,
$$
such that $ab-pq=1$. Conversely, it  can be shown that, given a Egorov system with a pair of conservation laws satisfying the property $ab-pq=1$, one can effectively reconstruct the first two  commuting flows in (\ref{nP1}) and, thus,  obtain a hydrodynamic reduction of  
(\ref{P1}). Therefore, the description of hydrodynamic reductions is reduced to the classification of Egorov's systems with a pair of conservation laws satisfying $ab-pq=1$. Although this problem seems to be complicated in general, explicit examples can be constructed by studying particular  Egorov's systems. As demonstrated in Examples 2 and 3, one and the same Egorov's system may possess different pairs of conservation laws satisfying the requirement $ab-pq=1$. It seems to be an interesting problem to classify such systems in general.

In  the two-component case, equations (\ref{nP2}) can be rewritten
in the new independent variables $a$ and $q$, assuming the form of a non-homogeneous system for $\mu^1, \mu^2$ and $w^1, w^2$:
\begin{equation}
\begin{array}{c}
\partial_a\mu^1+\mu^2\partial_q\mu^1=
\frac{w^1(\mu^1-\mu^2)^2}{w^1(a\mu^2-q)-w^2(a\mu^1-q)}, ~~~
\partial_a\mu^2+\mu^1\partial_q\mu^2=
\frac{w^2(\mu^1-\mu^2)^2}{w^2(a\mu^1-q)-w^1(a\mu^2-q)}, \\
\ \\
\partial_aw^1+\mu^2\partial_qw^1=
\frac{w^1(w^1-w^2)(\mu^1-\mu^2)}{w^1(a\mu^2-q)-w^2(a\mu^1-q)}, ~~~
\partial_aw^2+\mu^1\partial_qw^2=
\frac{w^2(w^1-w^2)(\mu^1-\mu^2)}{w^2(a\mu^1-q)-w^1(a\mu^2-q)}.
\label{GibTsa}
\end{array}
\end{equation}
Equations (\ref{GibTsa}) are a direct analog of the Gibbons-Tsarev system \cite{GibTsa96, GibTsa99} governing two-component hydrodynamic reductions of the Benney moment equations.

\section{One-component reductions} 

One-component reductions of (\ref{P3}) are of the form $a=a(R)$, $b=b(R)$,   $p=p(R)$,  $q=q(R)$, where $R(x, y, z, t)$ solves a triple of  hydrodynamic type equations 
\begin{equation}
R_t=v(R)\ R_x, ~~ R_y=w(R)\ R_x, ~~ R_z=\mu (R)\ R_x, 
\label{P4}
\end{equation}
which, in the one-component case, are automatically commuting. We recall that the general solution of (\ref{P4}) is given by the
implicit hodograph formula
\begin{equation}
f(R)=x+v(R)t+w(R)y+\mu (R)z 
\label{P5}
\end{equation}
where $f(R)$ is  arbitrary. One can readily verify that equations (\ref{P3}) 
imply $\mu={q'}/{a'}, ~~~ v={wp'}/{a'}$, so that the formula (\ref{P5}) assumes the form
\begin{equation}
F(R)=a'(R)x+q'(R)z+w(R)(p'(R)t+a'(R)y),
\label{P7}
\end{equation}
where $F=a'f$. Moreover, equations (\ref{P3}) also imply that $a, b, p, q$ are subject to the constraints
\begin{equation}
ab-pq=1, ~~~ a'b'-p'q'=0.
\label{P8}
\end{equation}
Thus, one-component reductions of the heavenly equation are defined by the formulae (\ref{P2}) 
where the functions $a(R), b(R), p(R), q(R)$ satisfy the restrictions (\ref{P8}), and $R(x, y, t, z)$ is implicitly defined  by (\ref{P7}). These solutions depend on 
four arbitrary functions of a single variable (indeed, $F, w$ and any two of $a, b, p, q$ can be arbitrary).

{\bf Remark.} As pointed to us  by Y. Nutku, one can slightly `relax' the scheme described above,  considering only two hydrodynamic equations
\begin{equation}
R_t=\nu(R)\ R_y, ~~ R_z=\mu (R)\ R_x, 
\label{P9}
\end{equation}
which, in the one-component situation, are automatically compatible. Indeed, the general solution of
the system (\ref{P9}) is given by the implicit formula
\begin{equation}
F(R, t+\nu(R)y, z+\mu(R)x)=0, 
\label{P10}
\end{equation}
where $F$ is an arbitrary function of three arguments. Then equations (\ref{P3}) for
$a(R), b(R), p(R)$ and $q(R)$ imply 
\begin{equation}
\nu=\frac{p'}{a'}, ~~~ \mu=\frac{q'}{a'},
\label{P11}
\end{equation}
along with equations (\ref{P8}). Choosing   $a(R), b(R), p(R), q(R)$ which satisfy   
equations (\ref{P8}) and defining $\nu, \mu$ by the formulae (\ref{P11}), we reconstruct the solution
$\Omega(x, y, z, t)$ of the heavenly equation by virtue of (\ref{P2}) where $R(x, y, z, t)$ is defined by the implicit relation (\ref{P10}). 

We emphasize, however, that such `relaxation' of the general scheme applies to one-component case only: one can show that two strictly hyperbolic 
hydrodynamic type systems
$$
R^i_t=\nu ^i(R)\ R^i_y, ~~ R^i_z=\mu^i (R)\ R^i_x,  ~~~~ i\geq 2,
$$
are compatible if and only if there exists a third system
$$
R^i_y=w ^i(R)\ R^i_x
$$
which commutes with both.

\bigskip

\section{Examples of two-component reductions} 

{\bf Example 1.} Let us consider the following triple of commuting systems,
\begin{equation}
R^1_t=4 e^{2R^2-\frac{2}{3}R^1}\ R^1_x,  ~~~ 
R^2_t=12 e^{2R^2-\frac{2}{3}R^1}\ R^2_x, 
\label{E1}
\end{equation}
\begin{equation}
R^1_y=\frac{1}{4} e^{-4R^2-\frac{4}{3}R^1}\ R^1_x, ~~~
R^2_y=-\frac{3}{4} e^{-4R^2-\frac{4}{3}R^1}\ R^2_x, 
\label{E2}
\end{equation}
and
\begin{equation}
R^1_z=- e^{-2(R^1+R^2)}\ R^1_x, ~~~
R^2_z= e^{-2(R^1+R^2)}\ R^2_x;
\label{E3}
\end{equation}
the corresponding metric
$$
 e^{2(R^1+R^2)}[ (dR^1)^2+ (dR^2)^2]
$$
clearly satisfies the Egorov property.
Notice that  these systems belong to the hierarchy of commuting flows of equations of isentropic gas dynamics with the polytropic constant $\gamma=1$. Introducing 
$$
a=\frac{1}{2\sqrt 3}e^{\frac{2}{3}R^1-2R^2}, ~~ q=\frac{1}{4\sqrt 3}e^{-\frac{4}{3}R^1-4R^2}, ~~ 
p=\frac{4}{\sqrt 3}e^{\frac{4}{3}R^1+4R^2}, ~~  b=\frac{8}{\sqrt 3}e^{-\frac{2}{3}R^1+2R^2},
$$
one can readily verify the identities
(\ref{P3}). Therefore, the corresponding solution $\Omega$ of the heavenly equation (\ref{P1}) is
given by the formulae
$$
\Omega_{xy}=\frac{1}{2\sqrt 3}e^{\frac{2}{3}R^1-2R^2} , ~~~ 
\Omega_{zt}=\frac{8}{\sqrt 3}e^{-\frac{2}{3}R^1+2R^2}, ~~~ 
\Omega_{xt}=\frac{4}{\sqrt 3}e^{\frac{4}{3}R^1+4R^2}, ~~~ 
\Omega_{zy}=\frac{1}{4\sqrt 3}e^{-\frac{4}{3}R^1-4R^2};
$$
notice that these solutions are uniquely characterized by the property $\Omega_{xy}\Omega_{zt}=4/3$,  $\Omega_{xt}\Omega_{zy}=1/3$.
In fact, one can explicitly calculate {\it all} second derivatives of $\Omega$ in terms of $R^1, \ R^2$
as follows:
$$
\begin{array}{c}
\Omega _{xx} =\frac{2}{3\sqrt{3}}e^{2R^{1}+2R^{2}}, ~~ \Omega
_{xt}=\frac{4}{\sqrt{3}}e^{\frac{4}{3}R^{1}+4R^{2}}, ~~ \Omega
_{xy}=\frac{1}{2\sqrt{3}}e^{\frac{2}{3}R^{1}-2R^{2}}, ~~ \Omega_{xz}=-\frac{4}{3\sqrt{3}}(R^{1}-R^{2}), \\
\ \\
\Omega_{tt}=\frac{32}{\sqrt{3}}e^{\frac{2}{3}R^{1}+6R^{2}}, ~~ \Omega
_{yt}=\frac{4}{\sqrt{3}}(\frac{1}{3}R^{1}-3R^{2}), ~~
\Omega _{zt} =\frac{8}{\sqrt{3}}e^{-\frac{2}{3}R^{1}+2R^{2}}, ~~ 
\Omega_{yy}=-\frac{1}{2\sqrt{3}}e^{-\frac{2}{3}R^{1}-6R^{2}}, \\
\ \\
\Omega _{yz}=\frac{1}{4\sqrt{3}}e^{-\frac{4}{3}R^{1}-4R^{2}}, ~~
\Omega _{zz} =-\frac{2}{3\sqrt{3}}e^{-2R^{1}-2R^{2}};
\end{array}
$$
the compatibility conditions are identically satisfied by virtue of (\ref{E1})-(\ref{E3}). Here the Riemann invariants $R^1(x, y, z, t)$ and   $R^2(x, y, z, t)$ are implicitly defined by the hodograph relations
$$
\begin{array}{c}
\lambda^1(R)=x+4 e^{2R^2-\frac{2}{3}R^1}\ t+\frac{1}{4} e^{-4R^2-\frac{4}{3}R^1}\ y-2 e^{ -2(R^1+R^2)} \ z, \\
\ \\
\lambda^2(R)=x+12 e^{2R^2-\frac{2}{3}R^1}\ t-\frac{3}{4} e^{-4R^2-\frac{4}{3}R^1}\ y+2 e^{ -2(R^1+R^2)} \ z,
\end{array}
$$
where $\lambda^1(R),\   \lambda^2(R)$ are the characteristic speeds of commuting flows of
systems (\ref{E1}) - (\ref{E3})  satisfying the linear equations
\begin{equation}
\frac{\partial_2\lambda^1}{\lambda^2-\lambda^1}=1, ~~~~
\frac{\partial_1\lambda^2}{\lambda^1-\lambda^2}=1.
\label{comm}
\end{equation}
The general solution of these equations  can be represented in the form
$$
\lambda^1=-\partial_1(w e^{-R^1-R^2}), ~~~
\lambda^2=\partial_2(w e^{-R^1-R^2}),
$$
where the function $w$ solves the linear equation $\partial_1\partial_2w=w$. 

{\bf Example 2.} Modifying Example 1, let us consider the following triple of commuting systems of hydrodynamic type:
\begin{equation}
\begin{array}{c}
R^1_t=e^{-(R^1+R^2)}(\cos (R^2-R^1)-\sin (R^2-R^1))\ R^1_x,  \\
R^2_t=-e^{-(R^1+R^2)}(\cos (R^2-R^1)+\sin (R^2-R^1))\ R^2_x, 
\end{array}
\label{E4}
\end{equation}
\begin{equation}
\begin{array}{c}
R^1_y= e^{-(R^1+R^2)}(\cos (R^2-R^1)+\sin (R^2-R^1))\ R^1_x, \\
R^2_y= e^{-(R^1+R^2)}(\cos (R^2-R^1)-\sin (R^2-R^1))\ R^2_x, 
\end{array}
\label{E5}
\end{equation}
and
\begin{equation}
R^1_z=- e^{-2(R^1+R^2)}\ R^1_x, ~~~
R^2_z= e^{-2(R^1+R^2)}\ R^2_x.
\label{E6}
\end{equation}
Since the $z$-flow  (\ref{E6}) coincides with the $z$-flow (\ref{E3}) from Example 1,  the corresponding Egorov  metric will also be the same.  Introducing 
$$
\begin{array}{c}
a= e^{R^1+R^2}\cos (R^2-R^1), ~~ q= e^{-(R^1+R^2)}\sin (R^2-R^1), \\
\ \\
p= -e^{R^1+R^2}\sin (R^2-R^1), ~~  b= e^{-(R^1+R^2)}\cos (R^2-R^1),
\end{array}
$$
one can readily verify the identities
(\ref{P3}). The corresponding solution $\Omega$  is
given by the formulae
\begin{equation}
\begin{array}{c}
\Omega _{xx} =\frac{1}{2}e^{2(R^{1}+R^{2})}, ~~ 
\Omega_{xt}=-e^{R^{1}+R^{2}}\sin (R^{2}-R^{1}), ~~ 
\Omega_{xy}=e^{R^{1}+R^{2}}\cos (R^{2}-R^{1}), \\
\ \\
\Omega _{xz} =R^{2}-R^{1}, ~~ 
\Omega_{tt}=R^{1}+R^{2}+\sin ^{2}(R^{2}-R^{1}), ~~ 
\Omega _{yt}=-\sin(R^{2}-R^{1})\cos (R^{2}-R^{1}), \\
\ \\
\Omega _{zt} =e^{-(R^{1}+R^{2})}\cos (R^{2}-R^{1}), ~~
\Omega _{yy}=R^{1}+R^{2}-\sin ^{2}(R^{2}-R^{1}), \\
\ \\
\Omega _{yz} =e^{-(R^{1}+R^{2})}\sin (R^{2}-R^{1}), ~~
 \Omega_{zz}=-\frac{1}{2}e^{-2(R^{1}+R^{2})}.
\end{array}
\label{Omega0}
\end{equation}
Here the Riemann invariants $R^1(x, y, z, t)$ and   $R^2(x, y, z, t)$ are implicitly defined by the hodograph relations
$$
\begin{array}{c}
\lambda^1(R)=x+e^{-(R^1+R^2)}(\cos (R^2-R^1)-\sin (R^2-R^1))\ t +\\
e^{-(R^1+R^2)}(\cos (R^2-R^1)+ \sin (R^2-R^1))\ y-2 e^{ -2(R^1+R^2)} \ z, \\
\ \\
\lambda^2(R)=x-e^{-(R^1+R^2)}(\cos (R^2-R^1)+\sin (R^2-R^1))\ t + \\
e^{-(R^1+R^2)}(\cos (R^2-R^1)-\sin (R^2-R^1))\ y+2 e^{ -2(R^1+R^2)} \ z,
\end{array}
$$
where $\lambda^1(R),\   \lambda^2(R)$ are the characteristic speeds of commuting flows of
systems (\ref{E4}) - (\ref{E6}) which satisfy the linear equations (\ref{comm}) from example 1
(indeed, the flows (\ref{E1}), (\ref{E2}), (\ref{E3}), (\ref{E4}), (\ref{E5}) commute with each other).
In the new dependent variables 
$$
v=R^2-R^1, ~~~ \varphi =e^{R^1+R^2},
$$
equations (\ref{E4})-(\ref{E6}) take the conservative form
\begin{equation}
[\varphi ^2/2]_t=[-\varphi  \sin v]_x, ~~~ [v]_t=[\cos v/\varphi ]_x,
\label{E7}
\end{equation}
\begin{equation}
[\varphi ^2/2]_y=[\varphi \cos v]_x, ~~~ [v]_y=[\sin v/\varphi]_x,
\label{E8}
\end{equation}
and
\begin{equation}
[\varphi ^2/2]_z=[v]_x, ~~~ [v]_z=[-1/(2 \varphi ^2)]_x,
\label{E9}
\end{equation}
respectively. Written in terms of $u=\varphi^2$, equations (\ref{E9}) take the form
of the nonlinear elasticity equation $u_{zz}+(1/u)_{xx}=0$. The corresponding equations
(\ref{Omega0}) reduce to
$$
\begin{array}{c}
\Omega _{xx} =\frac{1}{2}\varphi ^{2}, ~~ \Omega
_{xt}=-\varphi \sin \upsilon, ~~ \Omega _{xy}=\varphi \cos
\upsilon, ~~ \Omega _{xz}=\upsilon,  \\
\ \\
\Omega _{tt} =\ln \varphi +\sin ^{2}\upsilon, ~~ \Omega
_{yt}=\sin \upsilon \cos \upsilon, ~~ \Omega _{zt}=
{\cos \upsilon }/{\varphi }, ~~ 
\Omega _{yy} =\ln \varphi -\sin ^{2}\upsilon, \\
\ \\
\Omega _{yz}={\sin \upsilon }/{\varphi }, ~~
\Omega _{zz}=-{1}/(2\varphi ^{2}).
\end{array}
$$

{\bf Example 3.} The elliptic version of equations (\ref{E7})-(\ref{E9}) is
\begin{equation}
[\varphi^2/2]_t=[\varphi \sinh v]_x, ~~~ [v]_t=[\cosh v/\varphi]_x,
\label{E10}
\end{equation}
\begin{equation}
[\varphi^2/2]_y=[\varphi \cosh v]_x, ~~~ [v]_y=[\sinh v/\varphi]_x,
\label{E11}
\end{equation}
and
\begin{equation}
[\varphi^2/2]_z=[v]_x, ~~~ [v]_z=[1/(2 \varphi^2)]_x,
\label{E12}
\end{equation}
respectively. Written in terms of $u=\varphi^2$,
equations (\ref{E12}) take the form of the nonlinear  elliptic equation $u_{zz}-(1/u)_{xx}=0$. The corresponding equations for $\Omega$ reduce to
$$
\begin{array}{c}
\Omega _{xx} ={\varphi ^{2}}/{2}, ~~ \Omega _{xt}=\varphi
\sinh \upsilon, ~~ \Omega _{xy}=\varphi \cosh \upsilon, ~~ \Omega _{xz}=\upsilon, ~~
\Omega _{tt} =\sinh ^{2}\upsilon -\ln \varphi, \\
\ \\
\Omega_{yt}=\sinh \upsilon \cosh \upsilon, ~~ \Omega _{zt}={\cosh \upsilon }/{\varphi }, ~~
\Omega _{yy} =\cosh ^{2}\upsilon +\ln \varphi, \\
\ \\
\Omega_{yz}={\sinh \upsilon }/{\varphi }, ~~ \Omega _{zz}={1}/(2\varphi ^{2}).
\end{array}
$$

\section{Examples of multi-component reductions}

{\bf Example 4.} The simplest possibility is to set $w^i=1$ in equations 
(\ref{nP2}). In this case  equations for $a$ simplify to $a\partial_i\partial_j a=2\partial_ia\partial_ja$, so that, up to a transformation $R^i\to f^i(R^i)$, the function $a$ can be
choosen in the form $a=1/(R^1+ ... + R^n)$.  Hence, equations (\ref{nP2}) become
\begin{equation}
\partial_iq=\mu^i\partial_ia, ~~~ 
\frac{\partial_j\mu^i}{\mu^j-\mu^i}=-\frac{1}{\sum R^k},
\label{w=1}
\end{equation}
while $v^i$ and  $p, \ b$ are defined by (\ref{v}) and (\ref{pb}).
We point out that the choice $w^i=1$ implies $\Omega=\Omega(\xi, z, t)$,  $\xi=x+y$, leading to the equation
\begin{equation}
\Omega_{\xi \xi}\Omega_{zt}-\Omega_{\xi t}\Omega_{\xi z}=1,
\label{mP1}
\end{equation}
which is a reduction of the heavenly equation by a self-dual symmetry  \cite{Boyer3, Finley}. Therefore,
equations (\ref{w=1}) (which are manifestly linear) describe hydrodynamic reductions of 
(\ref{mP1}). Choosing, for instance,
$$
\mu^i=c^i\sum R^k-\sum c^kR^k
$$
where $c^i$ are arbitrary constants, one can readily solve the remaining equations
(\ref{w=1}), (\ref{v}),  (\ref{pb}) for $q, \ v^i, \ p, \ b$, obtaining 
$$
v^i=\sum R^k/c^k-\frac{1}{c^i}\sum R^k
$$
and
\begin{equation}
\begin{array}{c}
a=\frac{1}{\sum R^k}, ~~~ q=-\frac{\sum c^kR^k}{\sum R^k}, \\
\ \\
p=\frac{\sum R^k/c^k}{\sum R^k}, ~~~
b=\sum R^k-\frac{(\sum c^kR^k)(\sum R^k/c^k)}{\sum R^k}.
\end{array}
\label{aqpb}
\end{equation}
Thus, we have constructed hydrodynamic reductions of the equation (\ref{mP1}):
\begin{equation}
\begin{array}{c}
R^i_t=\left ( \sum R^k/c^k- \frac{1}{c^i}\sum R^k \right )\ R^i_{\xi}, \\
\ \\
R^i_z=\left( c^i\sum R^k-\sum c^kR^k\right ) \ R^i_{\xi};
\end{array}
\label{xi}
\end{equation}
the associated  Egorov metric is
$$
ds^2=\frac{(dR^1)^2+ ... + (dR^n)^2}{(\sum R^i)^2}.
$$
The corresponding solution $\Omega(\xi, z, t)$ is given by the formulae
\begin{equation}
\begin{array}{c}
\Omega_{\xi \xi}=\frac{1}{\sum R^k}, ~~~ 
\Omega_{zt}=\sum R^k-\frac{(\sum c^kR^k)(\sum R^k/c^k)}{\sum R^k}, \\
\ \\
\Omega_{\xi t}=\frac{\sum R^k/c^k}{\sum R^k}, ~~~ 
\Omega_{\xi z}=-\frac{\sum c^kR^k}{\sum R^k}.
\end{array}
\label{O}
\end{equation}
Here the dependence of $R^i$ on $\xi, t, z$ is defined by the generalized hodograph relations
$$
\lambda^i(R)=\xi + \left ( \sum R^k/c^k -\frac{1}{c^i}\sum R^k\right )\ t+
\left( c^i\sum R^k-\sum c^kR^k\right )\ z,
$$
where $\lambda^i(R)$ are the characteristic speeds of commuting flows of systems (\ref{xi}),
that is, solutions of the linear system
$$
\frac{\partial_j\lambda^i}{\lambda^j-\lambda^i}=-\frac{1}{\sum R^k}.
$$
One can show that the general solution of this system is given by the formula
$$
\lambda^i(R)=f_i'(R^i)\sum R^k-\sum f_k(R^k),
$$
where $f_k(R^k)$ are $n$ arbitrary functions of a single argument. 

{\bf Remark.} Equations (\ref{xi}) can be obtained from the linear systems
$$
 R^i_{\tilde t}=-\frac{1}{c^i} \ R^i_{\tilde \xi}, ~~~ R^i_{\tilde z}=c^i \ R^i_{\tilde \xi}, 
$$
 by a reciprocal transformation
\begin{equation}
\begin{array}{c}
d\xi=(\sum R^k)\ d\tilde \xi -  (\sum R^k/c^k)\ d \tilde t+(\sum c^kR^k)\ d\tilde z,\\
\ \\
t=\tilde t, ~~~ z=\tilde z.
\end{array}
\label{recip}
\end{equation} 
Notice that, by virtue of (\ref{O}), the first equation (\ref{recip}) implies $\tilde \xi=\Omega_{\xi}$.
This reflects  the linearizability of the equation
(\ref{mP1}) by the contact transformation
$$
\tilde \xi=\Omega_{\xi}, ~~ \tilde t=t, ~~ \tilde z=z, ~~ \tilde \Omega={\xi }\Omega_{\xi}-\Omega, ~~
\tilde \Omega_{\tilde \xi}=\xi, ~~ \tilde \Omega_{\tilde t}=-\Omega_t, ~~ \tilde \Omega_{\tilde z}=
-\Omega_z,
$$
transforming (\ref{mP1}) to the linear wave equation
$$
\tilde \Omega_{\tilde \xi \tilde \xi}+\tilde \Omega_{\tilde t \tilde z}=0;
$$
see, e.g., \cite{Boyer3, Finley}.

{\bf Example 5.} Let us first recall the definition of the equations of associativity (the case of 4 primary fields), following \cite{Dubr}. One starts with a function $F(t^1, t^2, t^3, t^4)$ whose dependence on $t^1$ is specified as follows:
$$
F(t^1, t^2, t^3, t^4)=\frac{1}{2}(t^1)^2t^4+t^1t^2t^3+f(t^2, t^3, t^4).
$$
Introducing $c_{\alpha \beta \gamma}=\frac{\partial^3F}{\partial t^{\alpha} \partial t^{\beta} \partial
t^{\gamma}}$ and  $\eta_{\alpha \beta}=c_{1 \alpha \beta }$, defining the structure constants
$c_{\alpha \beta}^{ \gamma}=\eta^{\gamma \delta}c_{\delta \alpha \beta}$, and  requiring  the algebra with these structure constants to be associative, one arrives at the following system of third order PDEs for the function $f(t^2, t^3, t^4)$:
\begin{equation}
\begin{array}{c}
-2f_{234}-f_{233}f_{223}+f_{333}f_{222}=0, \\
\ \\
-f_{244}-f_{233}f_{224}+f_{334}f_{222}=0, \\
\ \\
-f_{344}-f_{223}f_{334}+f_{224}f_{333}=0, \\
\ \\
-2f_{234}f_{224}+f_{244}f_{223}+f_{344}f_{222}=0, \\
\ \\
f_{444}-f_{234}^2+f_{244}f_{233}-f_{334}f_{224}+f_{344}f_{223}=0, \\
\ \\
f_{333}f_{244}-2f_{334}f_{234}+f_{344}f_{233}=0; 
\end{array}
\label{ass}
\end{equation}
subscripts denote partial differentiation with respect to $t^2, t^3, t^4$. Following \cite{Dubr}, one introduces a triple of  hydrodynamic type systems
\begin{equation}
 t^{\beta}_t=c^{\beta}_{2\gamma}t^{\gamma}_x,  ~~~
 t^{\beta}_y=c^{\beta}_{3\gamma}t^{\gamma}_x,  ~~~ t^{\beta}_z=c^{\beta}_{4\gamma}t^{\gamma}_x,  ~~~ \beta, \gamma = 1, 2, 3, 4,
\label{eq}
\end{equation}
which commute with each other. Let us take 
$$
f(t^2, t^3, t^4)=\frac{(t^2)^2(t^3)^2}{2t^4}-t^4\ln t^3+t^4\ln t^4,
$$
which is a solution of (\ref{ass}). Introducing the notation
$t^1=u, \ t^2=a, \ t^3=p, \ t^4=r$, one can write  the flows (\ref{eq}) in the conservative form as follows:
\begin{equation}
\begin{array}{c}
r_{t} =\partial _{x}p, ~~   u_{t}=-\partial _{x}\frac{ap^{2}}{
r^{2}}, ~~ a_{t}=\partial _{x}[u+\frac{2ap}{r}], ~~ p_{t}=\partial _{x}\frac{p^{2}}{r}, \\
\ \\
r_{y} =\partial _{x}a, ~~ u_{y}=-\partial _{x}[\frac{1}{p}+
\frac{a^{2}p}{r^{2}}], ~~ a_{y}=\partial _{x}[\frac{a^{2}}{r}+
\frac{r}{p^{2}}], ~~ p_{y}=\partial _{x}[u+\frac{2ap}{r}], \\
\ \\
r_{z} =\partial _{x}u, ~~ u_{z}=\partial _{x}[\frac{1}{r}+
\frac{a^{2}p^{2}}{r^{3}}], ~~ a_{z}=-\partial _{x}[\frac{1}{p}+
\frac{a^{2}p}{r^{2}}], ~~ p_{z}=-\partial _{x}\frac{ap^{2}}{r^{2}};
\end{array}
\label{flows}
\end{equation}
(we point out that these equations are not hyperbolic).
One can show by a direct calculation that there exists a potential $\Omega$ whose second derivatives are given by the formulae
\begin{equation}
\begin{array}{c}
\Omega _{xx} =r, ~~ \Omega _{xt}=p, ~~ \Omega _{xy}=a, ~~
\Omega _{xz}=u, ~~ \Omega _{tt}=\frac{p^{2}}{r}, ~~ \Omega _{yt}=u+2\frac{ap}{r}\text{,} \\
\ \\
\Omega _{zt} =-\frac{ap^{2}}{r^{2}}, ~~ \Omega_{yy}=\frac{r}{
p^{2}}+\frac{a^{2}}{r}, ~~ \Omega _{yz}=-\frac{1}{p}(1+\frac{
a^{2}p^{2}}{r^{2}}), ~~ \Omega _{zz}=\frac{1}{r}(1+\frac{
a^{2}p^{2}}{r^{2}});
\end{array}
\label{second}
\end{equation}
clearly, $\Omega$ solves the equation (\ref{P1}).
Introducing the variables $\mu=p/r, \ \upsilon=u+ap/r$, the general solution of the flows (\ref{flows})
can be written in the implicit form
$$
\begin{array}{c}
z =a\Phi _{\upsilon \upsilon }-\Phi _{\mu \upsilon }+rG_{\upsilon }, \\
\ \\
y=\Phi _{\upsilon }+\mu \lbrack a\Phi _{\upsilon \upsilon }-\Phi
_{\mu \upsilon }+rG_{\upsilon }], \\
\ \\
t =\frac{a^{2}}{r}\Phi _{\upsilon \upsilon }+aG_{\upsilon }-\frac{1}{r}
\Phi _{\mu \mu }+G_{\mu }, \\
\ \\
x =G-\mu \lbrack \frac{a^{2}}{r}\Phi _{\upsilon
\upsilon }+aG_{\upsilon }-
\frac{1}{r}\Phi _{\mu \mu }+G_{\mu }],
\end{array}
$$
where  $\Phi (\upsilon $, $\mu )$ and $G(\upsilon $, $\mu )$ solve the equations
$$
\Phi _{\upsilon \upsilon }+\mu ^{2}\Phi _{\mu \mu }=0, ~~~
G_{\upsilon \upsilon }+\mu ^{2}G_{\mu \mu }=0.
$$
Notice that equations (\ref{second}) for the potential $\Omega$ can be  integrated twice,
resulting in the expression
$$
\begin{array}{c}
\Omega =\left(\frac{a^{2}}{r}\Phi _{\upsilon \upsilon }+aG_{\upsilon }-\frac{
\Phi _{\mu \mu }}{r} +G \right)(\upsilon \Phi _{\upsilon }-\Phi )- \\
(a\Phi _{\upsilon
\upsilon }+rG_{\upsilon }-\Phi _{\mu \upsilon })(\mu R+S)-R\Phi _{\upsilon }+H,
\end{array}
$$
where the functions $R, S, H$ are defined by the quadratures
$$
\begin{array}{c}
dR =G_{\mu }d\upsilon -\frac{G_{\upsilon }}{\mu ^{2}}d\mu, \\
\ \\
dS=(\frac{G_{\upsilon }}{\mu }-\upsilon G_{\mu })d\mu -(\mu G_{\mu
}+\upsilon G_{\upsilon })d\upsilon, \\
\ \\
dH =[\Phi _{\upsilon }G_{\mu }+\Phi _{\mu }G_{\upsilon }]d\upsilon +[\Phi
_{\mu }G_{\mu }-\frac{\Phi _{\upsilon }G_{\upsilon }}{\mu ^{2}}]d\mu .
\end{array}
$$
It is an interesting problem to find other solutions of the associativity equations which 
produce solutions of (\ref{P1}).

\section{Acknowledgements}
 
We thank R. Halburd, Y. Nutku, M. Sheftel  and I. Strachan for their interest and useful remarks.

\end{document}